%% file: dof_2d_gravity.tex
\documentclass[10pt, a4paper]{article}
\pdfoutput=1
\usepackage[utf8]{inputenc}
\usepackage[T1]{fontenc}
\usepackage{lmodern, textcomp}
\usepackage[british]{babel}
\usepackage{csquotes}

\usepackage{eurosym, xspace}
\usepackage[nottoc]{tocbibind}
\usepackage{graphicx, subcaption, float}
\usepackage{marginnote}
\usepackage{xparse}
\usepackage{authblk}

\usepackage[usenames, dvipsnames, svgnames, table]{xcolor}

\usepackage[backend=bibtex8, citestyle=numeric-comp, maxbibnames=99,
			sorting=none, sortcase=false, sortcites=true, giveninits=true]{biblatex}

\usepackage{amsmath, amsfonts, amssymb}
\usepackage{mathtools, cancel}
\usepackage{mathrsfs}
\usepackage{siunitx}
\usepackage{slashed, tensor}

\usepackage{maths_min}

\usepackage{hyperxmp}
\usepackage[pdftex, breaklinks=true, linktocpage,
	colorlinks=true, urlcolor=blue, linkcolor=blue, citecolor=red,
	pdfauthor={Harold Erbin}
]{hyperref}
\usepackage[all]{hypcap}

\DeclareUnicodeCharacter{00A0}{~}
\DeclareUnicodeCharacter{202F}{~}

\newcommand{\email}[1]{\href{mailto:#1}{\nolinkurl{#1}}}
\newcommand{\emailfoot}[1]{\thanks{\email{#1}}}

\newcounter{draftcommentcnt}
\NewDocumentCommand{\draftcomment}{s O{red} m}{%
	\def\margnote{\IfBooleanTF{#1}{\marginnote}{\marginpar}}%
	\stepcounter{draftcommentcnt}%
	\textcolor{#2}{#3}%
	\margnote{\textcolor{#2}{$\Leftarrow$ \arabic{draftcommentcnt}}}%
}

\graphicspath{{images/}}
\numberwithin{equation}{section}

\usepackage[sort&compress, english]{cleveref}

\usepackage[heightrounded,
	top=4cm, bottom=4cm, left=3.5cm, right=3.5cm,
]{geometry}

\input{arxiv.def}

\hypersetup{
	pdftitle={A short note on dynamics and degrees of freedom in 2d classical gravity},
	pdfkeywords={2d gravity, degrees of freedom},
}

\title{A short note on dynamics and degrees of freedom in $2d$ classical gravity}

\author[1]{Corinne de Lacroix\emailfoot{lacroix@lpt.ens.fr}}
\author[1]{Harold Erbin\emailfoot{erbin@lpt.ens.fr}}
\affil[1]{\textsc{Cnrs}, \textsc{Lptens}, École Normale Supérieure, F-75231 Paris, France}

\begin{document}
\maketitle

\begin{abstract}
We comment on some peculiarities of matter with and without Weyl invariance coupled to classical $2d$ Einstein--Hilbert gravity for several models, in particular, related to the counting of degrees of freedom and on the dynamics.
We find that theories where the matter action is Weyl invariant has generically more degrees of freedom than action without the invariance.
This follows from the Weyl invariance of the metric equations of motion independently of the invariance of the action.
Then, we study another set of models with scalar fields and show that solutions to the equations of motion are either trivial or inconsistent.
To our knowledge, these aspects of classical $2d$ gravity have not been put forward and can be interesting to be remembered when using it as a toy model for $4d$ gravity.
The goal of this note is also as a pedagogical exercise: our results follow from standard methods, but we emphasize more direct computations.
\end{abstract}

\newpage
\vspace{\stretch{1}}

\hrule
\tableofcontents
\bigskip
\hrule


\input{sections/introduction}
\input{sections/classical_gravity}
\input{sections/dynamics}
\input{sections/dofs}

\section*{Acknowledgments}

We would like to thank Costas Bachas, Tresa Bautista, Antoine Bourget and Atish Dabholkar for useful discussions.
We are particularly grateful to Adel Bilal for carefully reading the draft of the manuscript, and to an anonymous referee for useful feedback.

\appendix

\input{sections/quantum_gravity}
\input{sections/chiral_anomaly}


\printbibliography[heading=bibintoc]

\end{document}

%% file: sections/introduction.tex
\section{Introduction}

Two-dimensional gravity has been a toy model for four-dimensional gravity for several decades.
Indeed, computations are generally tractable and one can hope that gravity in all dimensions could share some general attributes.
Moreover, there are hints from various approaches that quantum gravity becomes effectively $2$-dimensional in some regimes~\cite{Carlip:2016:SpontaneousDimensionalReduction, Carlip:2017:DimensionDimensionalReduction}.
Finally, two-dimensional gravity appears in the worldsheet formulation of string theory which provides a second incentive to study it.

$2d$ classical gravity is generically absent from standard references because it is deemed to be trivial and one is more interested in the quantum effects -- in particular when the matter is conformal~\cite{Polyakov:1981:QuantumGeometryBosonic}, which is the natural stage for string theory.
Nonetheless, the classical theory presents some peculiarities that have not been discussed elsewhere (to our knowledge) and we believe that it is useful to be aware of those.
This also illustrates that one should be careful when counting degrees of freedom and relating them to symmetries.
The results presented in this paper follow straightforwardly from more general methods~\cite{Henneaux:1994:QuantizationGaugeSystems, Kaparulin:2013:ConsistentInteractionsInvolution}.
However, we present here a more direct approach: albeit less general, this allows a simpler and more intuitive discussion of the degrees of freedom, with aims to be a pedagogical exercise.

We first discuss is the dynamics of unitary matter with and without Weyl invariance (local rescalings of the metric) coupled to classical gravity.
For concreteness, we consider the case of $N$ scalar fields with canonical kinetic term and general potential.
We find that the dynamics is trivial -- the only classical solutions are the ones for constant fields -- or even that the equations of motion are inconsistent with each other.
This shows that the constraints imposed by gravity are very restrictive.

The second aspect concerns the counting of degrees of freedom for a large class of Lagrangians.
We show that there are \emph{less} degrees of freedom when the matter action is not invariant under Weyl transformations compared to the case where they are invariant.
This is contrary to the standard expectations when discussing degrees of freedom and symmetries: generically, additional gauge symmetries remove degrees of freedom instead of adding some.
The derivation is also interesting in itself because it shows that the equations of motion for the metric are Weyl invariant even if the action is not.

To conclude, $2d$ gravity displays some peculiarities not found in higher dimensions.
This point should be kept in mind when comparing two and four dimensions.
This leads to another remark on the relation between the classical theory and the semi-classical limit of the quantized theory.
One often blames theories for which the latter does not reproduce the former and that there should be an inconsistency in the theory.
However, this is exactly what happens in $2d$ gravity: the specific properties of the degrees of freedom counting and the dynamics imply a mismatch between the classical and semi-classical theories (and there is no doubt that quantized $2d$ gravity and its semi-classical limit are consistent~\cite{Polyakov:1981:QuantumGeometryBosonic}, even in cases where the classical dynamics is trivial).
This may raise some doubts that $2d$ gravity is a good toy model for $4d$ gravity.

The paper is organized as follows.
In \Cref{sec:classical}, we describe the coupling of conformal and non-conformal matter to classical two-dimensional gravity.
Then in \Cref{sec:dynamics}, we study the dynamics for a general model with scalar fields.
Finally, in \Cref{sec:dofs}, we count the degrees of freedom.
In \Cref{app:quantum-gravity}, we make the link with the quantum theory.
\Cref{app:chiral-anomaly} compares this discussion with the more familiar setting of a $\group{U}(1)$ chiral gauge theory.

%% file: sections/classical_gravity.tex
\section{Classical two-dimensional gravity}
\label{sec:classical}

In this section, we review some general aspects of (classical) $2d$ gravity coupled to some matter $\Psi$.

Let $\mc M$ be a $2$-dimensional space with metric $g_{\mu\nu}$ with signature $(-+)$ and whose coordinates are denoted by~$x^\mu$.
The total action of the matter fields $\Psi$ coupled to $2d$ gravity is
\begin{equation}
    \label{2dgrav:action:gravity+matter}
    S[g, \Psi]
        = S_{\text{grav}}[g] + S_m[g, \Psi]
\end{equation}
where $S_{\text{grav}}$ is the action for pure gravity.
In all this paper, we restrict our considerations to the case where the matter action $S_m[g, \Psi]$ is obtained from the action $S_m[\eta, \Psi]$ on flat space $g_{\mu\nu} = \eta_{\mu\nu}$ through minimal coupling.\footnotemark{}
\footnotetext{%
    For example, this excludes terms of the form $R f(\Psi)$, where $R$ is the Ricci scalar.
}%
We also assume that there is no gauge symmetries -- beyond the diffeomorphisms and possibly the Weyl symmetry -- or additional constraints between the fields after coupling to gravity.
The action is required to satisfy the following criteria: renormalizability, invariance under diffeomorphisms and no more than first-order derivatives.

Weyl transformations are local rescalings of the metric
\begin{equation}
    \label{2dgrav:sym:weyl}
    g_{\mu\nu}(x)
        = \e^{2\omega(x)} g'_{\mu\nu}(x),
\end{equation}
and will be a central element of the discussion.
In general, this transformation is not a symmetry of the action.
A necessary condition for an action to be Weyl invariant is to be invariant under conformal transformations in flat space~\cite{Polchinski:1988:ScaleConformalInvariance, Iorio:1997:WeylGaugingConformalInvariance} (see also~\cite{Karananas:2016:WeylVsConformal, Karananas:2016:WeylRicciGauging, Farnsworth:2017:WeylConformalInvariance}).
This condition is sufficient when the action is at most quadratic in the first derivative, which is the case for the models considered in this paper.
Hence, we assume that the action for conformal matter is Weyl invariant after minimal coupling to gravity.

The gravitational action is given by the sum of two terms,
\begin{equation}
    \label{2dgrav:action:grav}
    S_{\text{grav}}
        = S_{\text{EH}} + S_\mu,
\end{equation}
as it is well-known that only two invariants fulfil the above conditions.
The first piece $S_{\text{EH}}$ is the Einstein--Hilbert action
\begin{equation}
    \label{2dgrav:action:einstein-hilbert}
    S_{\text{EH}}[g]
        = \int \dd^2 x \sqrt{\abs{g}}\, R
        = 4\pi \chi,
    \qquad
    \chi = 2 - 2 h,
\end{equation}
which is a topological invariant in two dimensions and equal to the Euler number $\chi$, $h$ being the genus of the surface.
As a consequence, it is not dynamical (equivalently, the Einstein tensor is identically zero) and it can be ignored as long as one is not interested in topological properties (which we are not).
This action is also invariant under Weyl transformations \eqref{2dgrav:sym:weyl}.

The second term is the cosmological constant:\footnotemark{}
\footnotetext{%
    The cosmological constant can be either positive or negative.
    In the current convention, $\mu > 0$ and $\mu < 0$ correspond respectively to anti-de Sitter and de Sitter.
}%
\begin{equation}
    \label{2dgrav:action:cosmological}
    S_\mu[g]
        = \mu \int \dd^2 x \sqrt{\abs{g}}
        = \mu\, A[g],
\end{equation}
where $A$ is the area of $\mc M$ associated to the metric $g$.
It is not invariant under Weyl transformations \eqref{2dgrav:sym:weyl}.
The presence of this term has dramatic consequences in two dimensions, as will be exemplified in \Cref{sec:dofs,sec:dynamics}.
In particular, in the case of pure gravity, the equation of motion reduces to $\mu = 0$ which has no solution because $\mu$ is a fixed parameter of the model.

The classical equations of motion are given by varying the full action \eqref{2dgrav:action:gravity+matter} with respect to $g_{\mu\nu}$ and $\Psi$:
\begin{equation}
    \frac{\var S}{\var g^{\mu\nu}}
        = 0,
    \qquad
    \frac{\var S}{\var \Psi}
        = 0.
\end{equation}
Without specifying the action for the matter, it is not possible to go further with the second equation.
Nonetheless, the first equation dictates a lot of properties of the system and is also responsible for its subtleties.

The energy--momentum tensors associated with $S$ and $S_m$ are defined by
\begin{equation}
    \label{2dgrav:eq:energy-tensor}
    T_{\mu\nu}
        = - \frac{4\pi}{\sqrt{\abs{g}}} \,
            \frac{\var S}{\var g^{\mu\nu}}, \qquad
    T^{(m)}_{\mu\nu}
        = - \frac{4\pi}{\sqrt{\abs{g}}}
            \frac{\var S_m}{\var g^{\mu\nu}}.
\end{equation}
Then, the metric equation of motion implies that the total energy--momentum tensor \eqref{2dgrav:eq:energy-tensor} vanishes:
\begin{equation}
    \label{2dgrav:eom:gravity+matter:g}
    T_{\mu\nu}
        = T^{(m)}_{\mu\nu} + 2\pi \mu\, g_{\mu\nu}
        = 0.
\end{equation}
These three independent equations provide constraints on the metric and matter fields (see \Cref{sec:dofs} for details on the counting of degrees of freedom).
It is convenient to decompose the energy--momentum tensor into its trace $T$ and its traceless components $\bar T_{\mu\nu}$:
\begin{equation}
    \bar T_{\mu\nu}
        := T_{\mu\nu} - \frac{1}{2} \; T\, g_{\mu\nu},
    \qquad
    T
        := g^{\mu\nu} T_{\mu\nu}.
\end{equation}
In terms of these variables, the equation of motion \eqref{2dgrav:eom:gravity+matter:g} reads
\begin{equation}
    \label{2dgrav:eom:gravity+matter:g:Tbar+trace}
    T = T^{(m)} + 4\pi \mu
        = 0, \qquad
    \bar T_{\mu\nu} = \bar T^{(m)}_{\mu\nu}
        = 0.
\end{equation}
An obvious advantage is the decoupling of the cosmological constant from the traceless tensor.
More generally, any matter potential that does not depend on the metric does not appear in $\bar T_{\mu\nu}$.

If $S_m[\eta, \Psi]$ is conformally invariant, then the action $S_m[g, \Psi]$ is Weyl invariant.
In this case, its energy--momentum tensor \eqref{2dgrav:eq:energy-tensor} is traceless $T^{(m)} = 0$.
Then, the trace of $T_{\mu\nu}$ reduces to the cosmological constant and \eqref{2dgrav:eom:gravity+matter:g:Tbar+trace} gives the equation of motion $\mu = 0$.
This equation has no solution since $\mu$ is fixed.
Hence, solutions for conformally invariant matter coupled to gravity exist only in the absence of a cosmological constant.
The equation of motion without cosmological constant is simply $T^{(m)}_{\mu\nu} = 0$, which provides two independent equations because it is symmetric and traceless.

The next section will generalize this analysis to the case of $N$ scalar fields with a general potential.
Then, we will describe the degrees of freedom for a more general class of models in \Cref{sec:dofs}.

%% file: sections/dynamics.tex
\section{Dynamics of unitary matter}
\label{sec:dynamics}

The goal of this section is to show that a large class of models of unitary matter coupled to gravity has no dynamics.\footnotemark{}
\footnotetext{%
    By unitary, we mean that the Hamiltonian is positive-definite (by analogy with CFT).
}%
We will illustrate this using $N$ scalar fields $X_i$ with a generic potential $V(X_i)$ (independent of the metric).

Weyl invariance of the matter action is possible only if $V = 0$.
In this case, it is well known that the energy--momentum tensor of free scalar fields can be rewritten as a sum of squares, which implies that each term vanishes independently.
This implies that classical solutions exist only for constant fields.
When the action is not Weyl invariant, the cosmological constant and the potential may provide a negative term in $T_{\mu\nu}$.
However, the decomposition \eqref{2dgrav:eom:gravity+matter:g:Tbar+trace} show that the equations in both cases can be brought into a similar form.
This indicates that the same behaviour could be expected; in fact, we will find that it is even worse.

We consider the following action:
\begin{equation}
    S_m
        = - \frac{1}{4\pi} \int \dd^2 x \sqrt{\abs{g}}\,
            \Big( g^{\mu\nu} \pd_\mu X_i \pd_\nu X_i + V(X_i) \Big),
    \qquad
    i = 1, \ldots, N,
\end{equation}
where the sum over the index $i$ is implicit and some of the $X_i$ may not appear in the potential.
This potential does not contain any constant term, which would just correspond to a shift of the cosmological constant, and it also does not contain the metric.
The equations of motion for the metric and the scalar fields are
\begin{equation}
    \label{2dgrav:eom:unitary-model}
    T_{\mu\nu}
        = 0,
    \qquad
    - \lap X_i + \frac{1}{2} \frac{\pd V}{\pd X_i}
        = 0,
\end{equation}
where $\lap$ is the curved space Laplacian for the metric $g$
\begin{equation}
    \label{conv:eq:laplacian}
    \lap
        = g^{\mu\nu} \grad_\mu \grad_\nu
        = \frac{1}{\sqrt{\abs{g}}}\, \big( \pd_\mu \sqrt{\abs{g}} g^{\mu\nu} \pd_\nu \big).
\end{equation}

Then, the trace and the traceless parts \eqref{2dgrav:eom:gravity+matter:g:Tbar+trace} of the energy--momentum tensor read
\begin{equation}
    \label{2dgrav:eom:unitary-trace-Tbar}
    T
        = - V + 4\pi \mu, \qquad
    \bar T_{\mu\nu}
        = \pd_\mu X_i \pd_\nu X_i
            - \frac{1}{2} g_{\mu\nu} (g^{\alpha\beta} \pd_\alpha X_i \pd_\beta X_i).
\end{equation}
One can see that $\bar T_{\mu\nu}$ is exactly the energy--momentum tensor of $N$ free scalar fields, and one can use the usual strategy to solve the associated equation.

First, one uses the diffeomorphisms to write the metric in the flat conformal gauge:
\begin{equation}
    g_{\mu\nu}
        = \e^{2\phi} \eta_{\mu\nu}.
\end{equation}
Writing explicitly the components of $\bar T_{\mu\nu}$, one arrives at the equation
\begin{equation}
    2 (\bar T_{00} \pm \bar T_{01})
        = (\pd_0 X_i \pm \pd_1 X_i)^2
        = 0.
\end{equation}
Since this is a sum of squares, the only solution is the one where all terms vanish independently:
\begin{equation}
    (\pd_0 \pm \pd_1) X_i
        = 0
    \quad \Longrightarrow \quad
    \pd_\mu X_i
        = 0
    \quad \Longrightarrow \quad
    X_i
        = X_i^0
        = \cst.
\end{equation}
In the conformal gauge, the matter equation \eqref{2dgrav:eom:unitary-model} and the trace equation give the following constraints on the values of $X^i_0$:
\begin{equation}
    \label{2dgrav:eq:unitary:constraint-X0}
    \frac{\pd V}{\pd X_i}(X_i^0)
        = 0,
    \qquad
    V(X_i^0)
        = 4 \pi \mu.
\end{equation}
It can be noted that no equation allows to fix the value of $\phi$.

The previous computation shows that unitary scalar fields coupled to $2d$ gravity do not have any dynamics since the solutions of the equations of motion is given by the trivial solution $X_i = \cst$ (with some constraints), even in the presence of interactions (if the potential does not contain the metric).\footnotemark{}
\footnotetext{%
    Note that this argument does not apply to the $X^\mu$ fields of bosonic string theory since $X^0$ is a (non-unitary) timelike scalar.
}%
The constraints provided by the vanishing of the energy--momentum tensor are too strong and kill all the dynamics (this is not the case for matter without gravity, should the fixed background be curved or not).

Finally, to give a specific example and to stress the difference between the Weyl invariant and non-invariant cases, we consider free massive scalar fields:
\begin{equation}
    V(X_i) = \sum_i m_i^2 X_i^2,
\end{equation}
where some of the masses can vanish, $m_i^2 \ge 0$.
The first equation \eqref{2dgrav:eq:unitary:constraint-X0} yields the conditions:
\begin{equation}
    m_i^2 X^0_i
        = 0
    \quad \Longrightarrow \quad
    X^0_i = 0,
    \qquad
    \forall m_i \neq 0.
\end{equation}
Introducing this into the second equation of \eqref{2dgrav:eq:unitary:constraint-X0} gives $\mu = 0$, which is incompatible with a non-vanishing cosmological constant.\footnotemark{}
\footnotetext{%
    Solutions can exist if at least one mass-squared is negative, but then the Hamiltonian is not positive-definite.
}%
This means that the dynamics is even more constrained when the matter is not Weyl invariant since there exist models which do not even have a solution.\footnotemark{}
\footnotetext{%
    Excluding the cosmological constant term when the theory is not Weyl invariant is a form of fine-tuning since it is not forbidden by symmetry.
}%

Interestingly, path integral computations lead to divergences which can be absorbed by renormalizing the cosmological constant.
Hence, consistency of the quantum theory requires inconsistency of the classical theory, at least in some cases (free massless and massive scalar fields).

%% file: sections/dofs.tex
\section{Degrees of freedom}
\label{sec:dofs}

In this section, we provide a counting of the degrees of freedom for a general Lagrangian at most linear in the inverse metric.
In this case, the equations of motion for the metric are Weyl invariant even if the action is not, as we will show below.
As a consequence, there are more constraints \emph{without} the Weyl symmetry than with it.
These results can also be derived from other methods, such as the one from~\cite{Kaparulin:2013:ConsistentInteractionsInvolution} or using a Hamiltonian analysis~\cite{Henneaux:1994:QuantizationGaugeSystems}.

Consider first pure gravity.
In two dimensions, the naive counting of on-shell degrees of freedom gives $-1$ because the system is over-constrained.
In general, a gauge invariance leads to two constraints, one from the gauge fixing condition and one from the equation of motion ensuring that the first one is preserved in time.
We have seen in \Cref{sec:classical} seen that pure gravity with a cosmological constant is inconsistent.
Setting $\mu = 0$ renders the theory Weyl invariant, which allows fixing the last component of the metric and there is no on-shell degree of freedom.

When matter is present, gravity reduces the number of degrees of freedom in $2$ dimensions.
We discuss now the case of a general set of matter fields $\Psi$ which have a total of $N$ on-shell degrees of freedom \emph{before} coupling to gravity.
The counting may seem useless given the absence of dynamics discussed in \Cref{sec:dynamics}, but it applies to more general situations (such as non-unitary matter).
Moreover, one can imagine generalizing the ideas discussed here by relaxing some assumptions -- we leave this for future work.

Since the precise counting is sensitive to the form of the action we restrict our attention to the simpler case when the matter Lagrangian is linear in the inverse metric:\footnotemark{}
\footnotetext{%
    This implies that covariant derivatives must reduce to simple derivatives: this is automatic for scalar fields, and this is also true for the kinetic term of Majorana fermions~\cite{Polyakov:1981:QuantumGeometryFermionic}.
}%
\begin{equation}
    S_m
        = \frac{1}{2\pi} \int \dd^2 x\, \sqrt{\abs{g}}\, \mc L,
    \qquad
    \mc L
        = - \frac{1}{2} \big( g^{\mu\nu} \mc L_{\mu\nu}(\Psi)
            + V(\Psi) \big).
\end{equation}
The first term is Weyl invariant because the transformation of $g^{\mu\nu}$ cancels the one of $\sqrt{\abs{g}}$.
Remember that we assume $\mc L_{\mu\nu}$ and $V$ to be such that there are no further constraints on the matter when coupling to gravity.\footnotemark{}
\footnotetext{%
    \label{2dgrav:ft:model-null-vec}
    An example of a model constrained after introducing gravity is $N$ scalar  fields with $\mc L_{\mu\nu}(X) = \eta_{ij} \pd_\mu X_i \pd_\nu X_j$, where $\eta_{ij}$ has a least one minus sign, and such the potential is aligned along a null vector $n_i$ in the field space: $V = \eta_{ij} n_i X_j$.
    In this case, there is a relation between the equations of motion:
    \[
        \eta_{ij} n_i \, \frac{\delta S_m}{\delta X_j}
            = - \frac{\sqrt{\abs{g}}}{2\pi} \, \lap V
            = \frac{\sqrt{\abs{g}}}{2\pi} \, \lap T
            = - \frac{1}{2} \, \lap \left( g^{\mu\nu} \frac{\delta S_m}{\delta g^{\mu\nu}} \right),
    \]
    which leads to additional gauge symmetry and constraints.
    We thank an anonymous referee for pointing out this caveat.
}%

The traceless and trace components \eqref{2dgrav:eom:gravity+matter:g:Tbar+trace} of the energy--momentum tensor are
\begin{equation}
    \label{2dgrav:eom:linear:g}
    \bar T_{\mu\nu}
        = \mc L_{\mu\nu} - \frac{1}{2}\, g_{\mu\nu} \big( g^{\alpha\beta} \mc L_{\alpha\beta} \big),
    \qquad
    T
        = - V + 4\pi\mu.
\end{equation}
The traceless part of the tensor is the energy--momentum tensor corresponding to the action $S_m$ where all the parameters breaking the Weyl invariance have been set to zero.
As a consequence, it is clear in the conformal gauge that no dynamical component (the conformal factor) of the metric appears inside and the equation $\bar T_{\mu\nu} = 0$ gives two constraints on the matter.
The trace is also Weyl invariant since it does not contain the metric and it provides another constraint for the matter.
In total, the number of on-shell degrees of freedom is reduced by three, giving $N - 3$.
After using the diffeomorphisms, the metric has still one off-shell degree of freedom which should be fixed by the equations of motion of the matter fields since they are not Weyl invariant.
However, the conformal factor is not expected to come with derivative -- see, for example, the Laplacian \eqref{conv:eq:laplacian} --, such that there is no additional on-shell degree of freedom associated to gravity and the total number if $N - 3$.\footnotemark{}
\footnotetext{%
    The previous statements hold directly in the conformal gauge.
    It should be adapted in other gauges, but the final number of degrees of freedom should be the same in any gauge.
}%

For $N$ scalar fields, this result can also be understood in the string theory language.\footnotemark{}
\footnotetext{%
    We thank an anonymous referee for this description.
}%
In that case, the action corresponds to a non-linear sigma model with target spacetime $\mc X$, metric $G_{ij}(X)$ and a tachyon potential $V(X)$.
The trace equation \eqref{2dgrav:eom:gravity+matter:g:Tbar+trace} implies that the spacetime $\mc X$ is restricted to an hypersurface $\Sigma$ satisfying the condition $V(X) = 4\pi \mu$.
If the normal vector $\pd_i V$ is timelike (in terms of $G_{ij}$), then the surface $\Sigma$ is spacelike and there is no dynamics (according to \Cref{sec:dynamics}).
If $\pd_i V$ is spacelike, then the action effectively describes a string moving in $N - 1$ dimensions, such that the final number of degrees of freedom is $N - 3$, according to the standard derivation.
Finally, additional constraints can emerge if $\pd_i V$ is null.\footnotemark{}
\footnotetext{%
    For example, the model in \Cref{2dgrav:ft:model-null-vec} has $N - 4$ degrees of freedom.
}%

If the matter action is Weyl invariant, then the trace equation is removed and there is one constraint less, giving $N - 2$ on-shell degrees of freedom.
As in the case of pure gravity, the last off-shell metric component is fixed by a Weyl transformation.

The counting provided in this section gives an upper-bound on the number of degrees of freedom.
Depending on the model, the constraints may be so strong that they remove even more degrees of freedom.
An extreme case is exemplified in \Cref{sec:dynamics} where no degree of freedom remains.

The conclusion is that even if Weyl invariance is not a symmetry of the action, it is a symmetry of the equations of motion for $g_{\mu\nu}$.
This is similar to the electric-magnetic duality, except that the matter equations of motion are breaking this duality.
It might be surprising to have a system where adding a gauge symmetry \emph{increases} the number of degrees of freedom, whereas the usual lore is that a gauge symmetry describes a redundancy.
This behaviour is very peculiar to two-dimensional gravity because the latter provides constraints together with degrees of freedom.
The metric components act as Lagrange multipliers and the Weyl symmetry removes one of those, which implies that one less constraint will be imposed.

%% file: sections/quantum_gravity.tex
\section{Quantum two-dimensional gravity}
\label{app:quantum-gravity}

In this appendix, we make contact between the classical and quantum regimes of gravity for the counting of the degrees of freedom.

The metric becomes dynamical due to quantum effects.
It is convenient to write the metric in the conformal gauge using the diffeomorphisms:
\begin{equation}
    \label{2dgrav:eq:conformal-gauge}
    g = \e^{2\phi} g_0,
\end{equation}
where $\phi$ is the Liouville mode and $g_0$ is a fixed background metric.
In this gauge, the dynamics of the metric and the matter decouple and both sectors are mixed only by implementing the constraints and integrating over the moduli parameters of the surface.

An alternative interpretation is to see \eqref{2dgrav:eq:conformal-gauge} as a parametrization instead of a gauge fixing.
Indeed, $g_0$ is arbitrary and it is desirable to put all choices on an equal footing by introducing an invariance under diffeomorphisms of $g_{0\mu\nu}$ (background diffeomorphisms).

Since the physical metric is left invariant under the transformation
\begin{equation}
    \label{2dgrav:sym:emergent-weyl}
    g_0
        = \e^{2\omega} g'_0,
    \qquad
    \phi
        = \phi' - \omega,
\end{equation}
it means that the system presents an emergent Weyl symmetry which is necessary to ensure that there is only one off-shell degree of freedom (three from $g_0$ and one from $\phi$, minus two from diffeomorphisms and one from this emergent symmetry).
This emergent Weyl symmetry is not fundamental since it is very specific to the conformal gauge, at the opposite of the Weyl symmetry \eqref{2dgrav:sym:weyl}: this last symmetry (when it exists) can be used in any gauge and truly reduce the total number of off-shell degrees of freedom, whereas the emergent Weyl is here only not to spoil the counting due to the redundant notation.
In particular, any action in the conformal gauge should have this invariance to be background-independent.

The total partition function is
\begin{equation}
    \label{quant:partition:gravity}
    Z
        = \int \dd_g g_{\mu\nu}\, \e^{- S_\mu[g]} Z_m|g],
        \qquad
    Z_m[g]
        = \int \dd_g \Psi\, \e^{- S_m[g, \Psi]},
\end{equation}
where $Z_m$ is the partition function of the matter and the index $g$ on the measure indicates that it depends on the metric.
Upon parametrizing the metric in the conformal gauge \eqref{2dgrav:eq:conformal-gauge}, one writes the partition function as:
\begin{equation}
    Z
        = \int \dd_g \phi \,
            \e^{- S_\mu[g_0, \phi] - S_{\text{grav}}[g_0, \phi]} \,
            \Delta_{\text{FP}}[g] Z_m|g_0],
\end{equation}
where $\Delta_{\text{FP}}[g]$ is the Faddeev--Popov determinant from the gauge fixing; it will be ignored in the rest of the discussion since it is not relevant.
The gravitational action
\begin{equation}
    \label{quant:action:grav}
    S_{\text{grav}}[g_0, \phi]
        = - \ln \frac{Z_m[\e^{2\phi} g_0]}{Z_m[g_0]}
\end{equation}
is the Wess--Zumino effective action\footnotemark{} for the change of metric from $g$ to $g_0$ in the matter partition function.
\footnotetext{%
    A Wess--Zumino action is not an effective action: instead, it is the difference between the effective action of the matter evaluated in both metrics.
    However, they share the property that they don't depend on the matter fields since they have been integrated out.
}%
Typically, the leading terms are the Liouville action and the Mabuchi action~\cite{Ferrari:2012:GravitationalActionsTwo, Bilal:2017:2DQuantumGravity, deLacroix:2016:MabuchiSpectrumMinisuperspace, deLacroix:2017:MinisuperspaceComputationMabuchi}.
The total action is then
\begin{equation}
    \label{quant:eq:quant-action}
    S_*[g_0, \phi, \Psi] = S_\mu[g_0, \phi] + S_{\text{grav}}[g_0, \phi] + S_m[g_0, \Psi].
\end{equation}

When the matter is Weyl invariant, the gravitational action can be computed by parametrizing the metric as $g = \e^{2\omega} g_0$ and by integrating the trace of the quantum energy--momentum tensor over $\omega$ from $0$ to $\phi$.
The latter is given by the Weyl anomaly (quantities with a subscript $0$ are given in terms of the metric $g_0$)
\begin{equation}
    \mean{T^m}
        = \frac{2\pi}{\sqrt{g}} \frac{\var}{\var \omega} \ln Z_m[g]
        = - \frac{c_m}{12}\, R.
\end{equation}
Using the relation $R = R_0 - 2 \lap_0 \omega$ and integrating, the action \eqref{quant:action:grav} becomes the Liouville action:
\begin{equation}
    \label{quant:action:liouville}
    S_{\text{grav}}
        = - \frac{c_m}{24\pi} \int \dd^2 x \sqrt{g_0}
            \big(g_0^{\mu\nu} \pd_\mu \phi \pd_\nu \phi + R_0 \phi \big).
\end{equation}

Studying the action \eqref{quant:eq:quant-action} as a classical action with background metric $g_0$ gives information about the semi-classical properties of the theory.
In \Cref{sec:dynamics}, we have argued that classical $2d$ gravity with unitary matter is trivial for a certain class of models.
But, this does not imply that the quantum behaviour is also trivial and that forbidden classical systems have no quantum dynamics.
As a specific example of this fact, consider the Liouville theory: classically, the cosmological constant is forbidden, but it is necessary to include it to compute the path integral (in particular, to absorb divergences and ambiguities~\cite{DHoker:1988:GeometryStringPerturbation}).
Surprisingly, the semi-classical limit does not resemble the classical gravity one started from: the quantum effects are not negligible in this limit and render the dynamics non-trivial.

Invariance under the emergent Weyl symmetry \eqref{2dgrav:sym:emergent-weyl} links the trace of the energy--momentum tensor $T^*_{0\mu\nu}$ to the variation of the full action with respect to the Liouville mode:
\begin{equation}
    \label{quant:eq:trace-energy-tensor-eom-phi}
    T_0^*
        = - \frac{1}{2}\, \frac{\var S_*}{\var \phi}.
\end{equation}

Another requirement is the one of background independence: the full action \eqref{quant:eq:quant-action} should be independent of the gauge choice \eqref{2dgrav:eq:conformal-gauge}, i.e.\ it should not depend on the background metric $g_0$ or on the specific decomposition.
This is true, in particular, if the variation of the total action with respect to $g_0$ vanishes, implying that the full stress--energy tensor is zero:
\begin{equation}
    \label{quant:eom:vanishing-quantum-T}
    T^*_{0\mu\nu}
        = 0.
\end{equation}
From \eqref{quant:eq:trace-energy-tensor-eom-phi} the trace equation is automatically satisfied if $\phi$ satisfies its equation of motion: as a consequence, \eqref{quant:eom:vanishing-quantum-T} provides two constraints on the matter and this number does not depend on whether the matter is Weyl invariant or not.

Finally, the variations of \eqref{quant:eq:quant-action} give decoupled equations of motion for $\phi$ and $\Psi$.
As a conclusion, one sees that quantum effects have given dynamics to the Liouville field and there is a total of $N - 1$ degrees of freedom (matter plus gravity) since the matter equations of motion are not expected to decrease further this number.

\section{Analogy: four-dimensional gauge anomaly}
\label{app:chiral-anomaly}

There exists a full analogy between the description of $2d$ gravity and $4d$ chiral gauge theory.
For example, one finds the emergence of a gauge symmetry (respectively Weyl and $\group{U}(1)$) through the choice of a convenient parametrization and the appearance of a Wess--Zumino action.

%% file: sections/chiral_anomaly.tex
\subsection{Massive vector field}

Let's consider a massive vector field $\mc A_\mu$ (playing the role of $g_{\mu\nu}$) with Proca action in $d$ dimensions:
\begin{equation}
    \label{chiral:action:proca}
    S_A
        = - \int \dd^d x \left(
            \frac{1}{4}\, \mc F_{\mu\nu} \mc F^{\mu\nu}
            + \frac{m^2}{2} \mc A_\mu \mc A^\mu
            \right).
\end{equation}
If $m^2 = 0$, then it enjoys a $\group{U}(1)_g$ gauge symmetry
\begin{equation}
    \label{chiral:sym:U1}
    \mc A_\mu
        = \mc A'_\mu + \pd_\mu \alpha,
\end{equation}
which reduces the $d$ components to $(d - 1)$ off-shell dofs, and furthermore to $(d - 2)$ on-shell dofs.
If $m^2 \neq 0$, then there is one on-shell constraint, such that there are only $(d-1)$ on-shell degrees of freedom.

In full similarity with $2d$ gravity, after coupling to matter fields, the matter action may or may not be invariant under the $\group{U}(1)_g$, even if $m^2 \neq 0$.

It is convenient to adopt another parametrization (called Stückelberg) for the vector field, where the spin $0$ component is separated from the spin $1$ component:
\begin{equation}
    \label{chiral:eq:param-A}
    \mc A_\mu
        = A_\mu + \pd_\mu a,
\end{equation}
where $A_\mu$ and $a$ (called the axion) play respectively the roles of $g_{0\mu\nu}$ and $\phi$ (from the point of view of Lorentz representations, the trace of $g_{\mu\nu}$ is similar to the divergence of $\mc A_\mu$), the only difference being that $A_\mu$ is dynamical.
In these variables, the action reads:
\begin{equation}
    \label{chiral:action:stuckelberg}
    S_A
        = - \int \dd^d x \left(
            \frac{1}{4}\, F_{\mu\nu} F^{\mu\nu}
            + \frac{m^2}{2} (A_\mu + \pd_\mu a) (A^\mu + \pd^\mu a)
            \right).
\end{equation}
In order to avoid introducing an additional degree of freedom ($A_\mu$ has $d$ components and $a$ has $1$), this system should be invariant under an emergent $\group{U}(1)_e$
\begin{equation}
    \label{chiral:sym:U1-emergent}
    A_\mu
        = A'_\mu + \pd_\mu \alpha,
    \qquad
    a
        = a' - \alpha.
\end{equation}
Note that this symmetry is not fundamental and is just a consequence of the parametrization \eqref{chiral:eq:param-A} that has been chosen.

\subsection{Effective action}

For the rest of this section, we focus on $d = 4$.
The action for the system is given by
\begin{equation}
    S[\mc A, \psi]
        = S_A[\mc A] + S_f[\mc A, \psi]
\end{equation}
where $S_A$ is Proca action \eqref{chiral:action:proca} and $S_f$ is the action for a chiral fermion $\psi$ (the dependence on the conjugate $\bar\psi$ is implicit everywhere)
\begin{equation}
    S_f
        = \int \dd^4 x \; i \bar \psi \slashed \dc \psi,
    \qquad
    \dc_\mu
        = \pd_\mu - i g \mc A_\mu.
\end{equation}
This model is discussed for example in~\cite{Preskill:1991:GaugeAnomaliesEffective}.
Since the fermion is chiral, it obeys the relations $\psi = \gamma_5 \psi = P_L \psi$ where $P_L$ is the projector on left-handed chirality.
The partition function of the system is
\begin{equation}
    \label{chiral:partition:matter}
    Z
        = \int \dd \mc A_\mu \,
            \e^{- S_A[\mc A]}\, Z_f[\mc A],
    \qquad
    Z_f[\mc A]
        = \int \dd\psi \,
            \e^{- S_f[\mc A, \psi]},
\end{equation}
where $Z_f$ is the fermion partition function.
The terms $S_A$ and $Z_f$ play respectively the roles of $S_\mu$ and $Z_m$ in \eqref{quant:partition:gravity}.

In terms of the parametrization \eqref{chiral:eq:param-A}, the partition function becomes (we ignore the ghost action since it decouples)
\begin{equation}
    Z
        = \int \dd A_\mu \dd a \,
            \e^{- S_A[A, a] - S_{\text{WZ}}[A, a]}\, Z_f[A],
\end{equation}
where the Wess--Zumino action is:
\begin{equation}
    S_{\text{WZ}}[A, a]
        = - \ln \frac{Z_f[A + \pd a]}{Z_f[A]}.
\end{equation}
The main difference with $2d$ gravity is that $A_\mu$ is dynamical and it does not drop from the path integral, but this does not modify the general argument.
Note that the gauge invariance \eqref{chiral:sym:U1-emergent} ensures that the same number of degrees of freedom is described as by \eqref{chiral:partition:matter}.

While the full action is not invariant $\group{U}(1)_g$ if $m^2 \neq 0$, the fermion action is invariant under the transformation $\psi = \e^{i g \alpha(x)} \psi'$ together with \eqref{chiral:sym:U1}, and the associated current $J^\mu = \bar\psi \gamma^\mu \psi$ is conserved classically.
On the other hand, there is a gauge anomaly due to the chirality of the theory and the current is not conserved quantum mechanically:
\begin{equation}
    \label{chiral:eq:quantum-anomaly}
    \pd_\mu \mean{J^\mu}
    = \frac{g^3}{48 \pi^2}\, \mc F_{\mu\nu} \widetilde{\mc F}^{\mu\nu}, \qquad
    \widetilde{\mc F}^{\mu\nu}
        = \frac{1}{2}\, \levi{^{\mu\nu\rho\sigma}} \mc F_{\rho\sigma}.
\end{equation}
This can be used to determine the $S_{\text{WZ}}$ in full similarity with the derivation of the Liouville action \eqref{quant:action:liouville}.
The quantum current is the variation of the effective action \eqref{chiral:partition:matter} with respect to the gauge field:
\begin{equation}
    \label{chiral:eq:quantum-current}
    \mean{J^\mu}
        = - \frac{\var}{\var \mc A_\mu} \ln Z_f[\mc A].
\end{equation}
Then, one can vary from $\mc A_\mu$ to $A_\mu$ continuously by parametrizing $\mc A_\mu = A_\mu + \pd_\mu \alpha$.
From the variation $\var \mc A_\mu = \pd_\mu \alpha$, the relation \eqref{chiral:eq:quantum-current} can be integrated from $\alpha = 0$ to $\alpha = a$ which results in (after an integration by part)
\begin{equation}
    S_{\text{WZ}}[A, a]
        = - \int \dd^4 x\, a\, \pd_\mu \mean{J^\mu}
        = - \frac{g^3}{48 \pi^2}\ \int \dd^4 x\, a \, F_{\mu\nu} \widetilde{F}^{\mu\nu},
\end{equation}
where the expression \eqref{chiral:eq:quantum-anomaly} and the identity $\mc F_{\mu\nu} = F_{\mu\nu}$ have been used.
This computation is possible because the mass term is outside the matter path integral which is effectively gauge invariant (classically).
In particular, the anomalous contribution arises from the fermion measure.
This mirrors $2d$ gravity where the cosmological constant lies outside the matter path integral.

The total action in the parametrization \eqref{chiral:eq:param-A} reads
\begin{equation}
    S[A, a, \psi]
        = S_A[A, a] + S_{\text{WZ}}[A, a] + S_f[A, \psi].
\end{equation}
This action should be invariant under the emergent $\group{U}(1)_e$; in particular this implies that the variation of the Wess--Zumino term $S_{\text{WZ}}$ is exactly the one necessary for cancelling the gauge anomaly of the fermion action \emph{in terms of the field} $A_\mu$ (the anomaly related to $\group{U}(1)_g$ is still present).
Note also that the fermions and the axion are not coupled to each other, in the same way, that the matter and the Liouville mode do not couple in $2d$ gravity.

Another point where the analysis differs from $2d$ gravity: the axion does not get its dynamics from the anomaly-generated term, but from the mass of $\mc A_\mu$ in \eqref{chiral:action:proca}.
One could then think that the axion is not dynamical if one starts with $m^2 = 0$.
However, such a mass term would be generated at $1$-loop from the cubic WZ vertex and the tree mass term is necessary to remove the divergence.
A similar story holds for the cosmological constant in $2d$ gravity.